\documentstyle[sprocl,epsfig]{article}

\begin{document}
\newcommand{\beq}{\begin{equation}}
\newcommand{\eeq}{\end{equation}}
\newcommand{\bea}{\begin{eqnarray}}
\newcommand{\eea}{\end{eqnarray}}
\title{DVCS  at HERA} 
\author{L.L.Frankfurt$^a$, A.Freund$^b$, M. Strikman$^b$}
\address{$^a$Physics Department, Tel Aviv University, Tel Aviv, Israel\\ 
$^b$Department of Physics, Penn State University\\
University Park, PA  16802, U.S.A.}

\maketitle\abstracts{We demonstrate that perturbative QCD allows one to 
calculate the absolute
cross section of diffractive exclusive production of photons
at large $Q^2$ at HERA, while the aligned jet model allows one to estimate 
the cross section for intermediate $Q^2 \sim 2 GeV^2$.  
 Furthermore, we find that the imaginary part of the amplitude for the 
production of real photons is larger than the imaginary part of the
corresponding  DIS amplitude by a factor of about $2$, leading to predictions 
of a significant 
counting rate for the current generation of experiments at HERA. We also find
a large azimuthal angle asymmetry in $ep$ scattering for
HERA kinematics which allows one to directly measure the real part of the 
DVCS amplitude and hence the nondiagonal parton distributions.}

Recent data from HERA has spurred great interest in exclusive or diffractive 
direct production of photons  in $e - p$ scattering (DVCS-
deeply virtual Compton scattering) as another source to obtain more 
information about the gluon distribution inside the proton for nonforward 
scattering. In recent years studies of diffractive vector meson production and
deeply virtual Compton scattering has greatly increased
our theoretical understanding about the gluon distribution in 
nonforward kinematics and how it compares to the gluon distribution in the 
forward direction, see list of references in Ref.\ \cite{1}.
We are interested in the production of a 
real photon compared to the inclusive DIS cross section. The
 exclusive process is nonforward 
in its nature, since the photon initiating the process is virtual 
($q^2<0$) and the final state photon is real, forcing a small but finite
momentum transfer to the target proton i.e forcing a nonforward kinematic 
situation as we would like. 

Similar to the case of deep inelastic scattering, in real photon
production it is possible to calculate within perturbative QCD
the $Q^2$ evolution of the amplitude but not its value at the normalization
point at $Q_0^2 \sim $ {\it few GeV$^2$}
where it is given by nonperturbative effects. 
Hence we start by discussing expectations for this region. It was demonstrated 
in \cite{FS88} that the aligned jet model coupled with the idea of 
color screening provides a reasonable semiquantitative description of
$F_{2N}(x \le 10^{-2}, Q_0^2)$. One can write $\sigma_{tot}(\gamma^*N)$
 using the Gribov dispersion representation \cite{Gribov} as \cite{FS88}:
\beq
\sigma_{tot}(\gamma^*N)= {\alpha \over 3 \pi}
 \int_{M_0^2}^{\infty}{\sigma_{tot}(``AJM''-N) R^{e^+e^-}(M^2)
M^2 {3 \left<k_{0~t}^2\right>\over M^2} \over (Q^2+M^2)^2}d M^2,
\label{AJMeq}
\eeq
Based on the logic of the local quark-hadron  duality 
(see e.g.\cite{FRS} and references therein) we take the
lower limit of integration $M_0^2 \sim 0.5 GeV^2 \le m_{\rho}^2$.
  In the case of real photon production the imaginary part of the 
amplitude for $t=0$ is obtained from Eq.\ \ref{AJMeq} by replacing one of the 
propagators by $1/M^2$. Using Eq.\ \ref{AJMeq} we find 
\beq
R \equiv {Im A(\gamma^*+N \to \gamma^* +N)_{t=0}
\over Im A(\gamma^*+N \to \gamma +N)_{t=0}}= {Q^2 \over Q^2+M_0^2}
\ln^{-1}(1+Q^2/m_0^2).
\label{ratiogam}
\eeq
For $M_0^2 \sim 0.4 \div 0.6~ GeV^2$  and 
$Q^2\approx 2-3~\mbox{GeV}^2$ Eq.\ref{ratiogam} leads to $R \approx 0.5$.
A similar value of $R$ has been found in \cite{FGS}.

The process of exclusive direct production of photons in 
first nontrivial order can be calculated  as the sum of  
hard contributions within the framework of QCD evolution equations 
\cite {Abramowicz}
and a soft contribution which we evaluated above within the aligned jet model. 
In order to calculate the imaginary part of the amplitude, we need to 
calculate the hard amplitude
as well as the gluon-nucleon scattering plus the 
soft aligned jet model contribution. 
We find the following  general expression for the 
imaginary part of the amplitude \cite{Abramowicz}:
\bea
ImA(x,Q^2,Q_0^2)&=& ImA(x,Q_0^2) + 4\pi^2\alpha\int_{Q_0^2}^{Q^2}
\frac{dQ'^2}{Q'^2}
\int^{1}_{x}\frac{dx_1}{x_1}[P_{qg}(x/x_1,\Delta /x_1)\nonumber\\
& &g(x_1,x_2,Q'^2) + P_{qq}(x/x_1,\Delta /x_1)q(x_1,x_2,Q'^2)],
\label{conv}
\eea
where $P_{qg}$ and $P_{qq}$ are the evolution kernels and are taken 
from Ref.\ \cite{1}.
For the  non-perturbative input, $Im A(x,Q_0^2)$
we will be able to use the previous aligned jet model analysis.
We now calculate the total imaginary part of the amplitude 
from Eq.\ \ref{conv} where we used CTEQ3L for the quark and gluon 
distributions. We neglect the $x_2$ dependence for the moment \cite{f5}.
The result for the ratio $R = ImA(\gamma^* + p \rightarrow \gamma^* + p)/
ImA(\gamma^* + p \rightarrow \gamma + p)$ is given in the $x$ 
range from $10^{-4}$ to $10^{-2}$ and for a $Q^2$ of $3.5,12$ and $45\; 
\mbox{GeV}^2$ relevant at HERA. According to our previous discussion we 
chose the initial distribution for the imaginary part of the DVCS amplitude 
to be twice that of the initial distribution for the imaginary part of the DIS
amplitude. We find $R$ to be between $0.551$, $0.573$ 
and $0.57$ for $x=10^{-4}$, $0.541$, $0.562$ and $0.557$ for $x=10^{-3}$ and 
$0.518$, $0.519$ and $0.505$ for $x=10^{-2}$ in the given $Q^2$ range. 
The ratio $R$ will approach $1/2$ as $Q^2$ is decreased to the 
nonperturbative scale since this is our aligned jet model estimate
The reason for the deviation from $R=1/2$ is due to the difference in the
evolution kernels.

As far as the complete amplitude at small $x$ is concerned, we can reconstruct
the real part via dispersion relations \cite{reim1,reim2} which to a 
very good 
approximation is given by:
\begin{equation}
\eta \equiv {Re A \over Im A}= \frac{\pi}{2}\frac{d\,\ln(x Im A)}{d\, 
\ln{1\over  x}}.
\label{reim}
\end{equation}
Therefore, our claims for the imaginary part of the 
amplitude  also hold for the whole amplitude at small $x$.

To check the feasibility of measuring a DVCS signal against the DIS 
background, 
we will be interested in the fractional number of DIS events
to DVCS events at HERA given by:
\bea
R_{\gamma} &=& \frac{\sigma(\gamma^* +p \rightarrow \gamma + p)}
{\sigma_{tot}(\gamma^*p)}\simeq
\frac{d\sigma (\gamma^* + p \rightarrow \gamma +p)}{dt}|_{t=0}\times 
\frac{1}{B}/\sigma_{tot}(\gamma^*p)\nonumber\\
&=&\frac{\pi \alpha}{4 R^2 Q^2 B}F_2(x,Q^2)(1 + \eta^2),
\label{events}
\eea
where $\eta= Re\,A/Im\,A\simeq 0.09-0.27$ for the 
given $Q^2$ range, is given by 
Eq.\ \ref{reim}. Note that only $d\sigma/dt(t=0)$ 
is calculable in QCD. The $t$ dependence is taken from the data fits to hard 
diffractive processes. 
We computed $R_{\gamma}$ for $x$ between
$10^{-4}$ and $10^{-2}$ and for a $Q^2$ of $2, 3.5, 12$ and $45\,\mbox{GeV}^2$ 
with the following results, where the numbers for $F_2$ were taken from
\cite{15}. We find $R_{\gamma}\simeq 1.1\times 10^{-3},\, 4.9\times 10^{-4}$ 
 at $x=10^{-4},\, 10^{-3}$ and $Q^2=2\mbox{GeV}^2$; $R_{\gamma}
\simeq 1.07\times 10^{-3},\, 9.3\times 10^{-4}$ 
  at $x=10^{-4},\, 10^{-3}$ and  $Q^2=3.5\mbox{GeV}^2$; 
$R_{\gamma}\simeq 4.5\times 10^{-4},\, 3.78\times 10^{-4}\, 
2.5\times 10^{-4}$ 
at $x=10^{-4},\, 10^{-3},\, 10^{-2}$ and $Q^2=12\mbox{GeV}^2$; and 
finally $R_{\gamma} \simeq 1.49\times 10^{-4},\, 1.04\times 10^{-4}$
at $x=10^{-3},\, 10^{-2}$ and $Q^2=45\mbox{GeV}^2$. 

\begin{figure}
\centerline{\epsfig{file=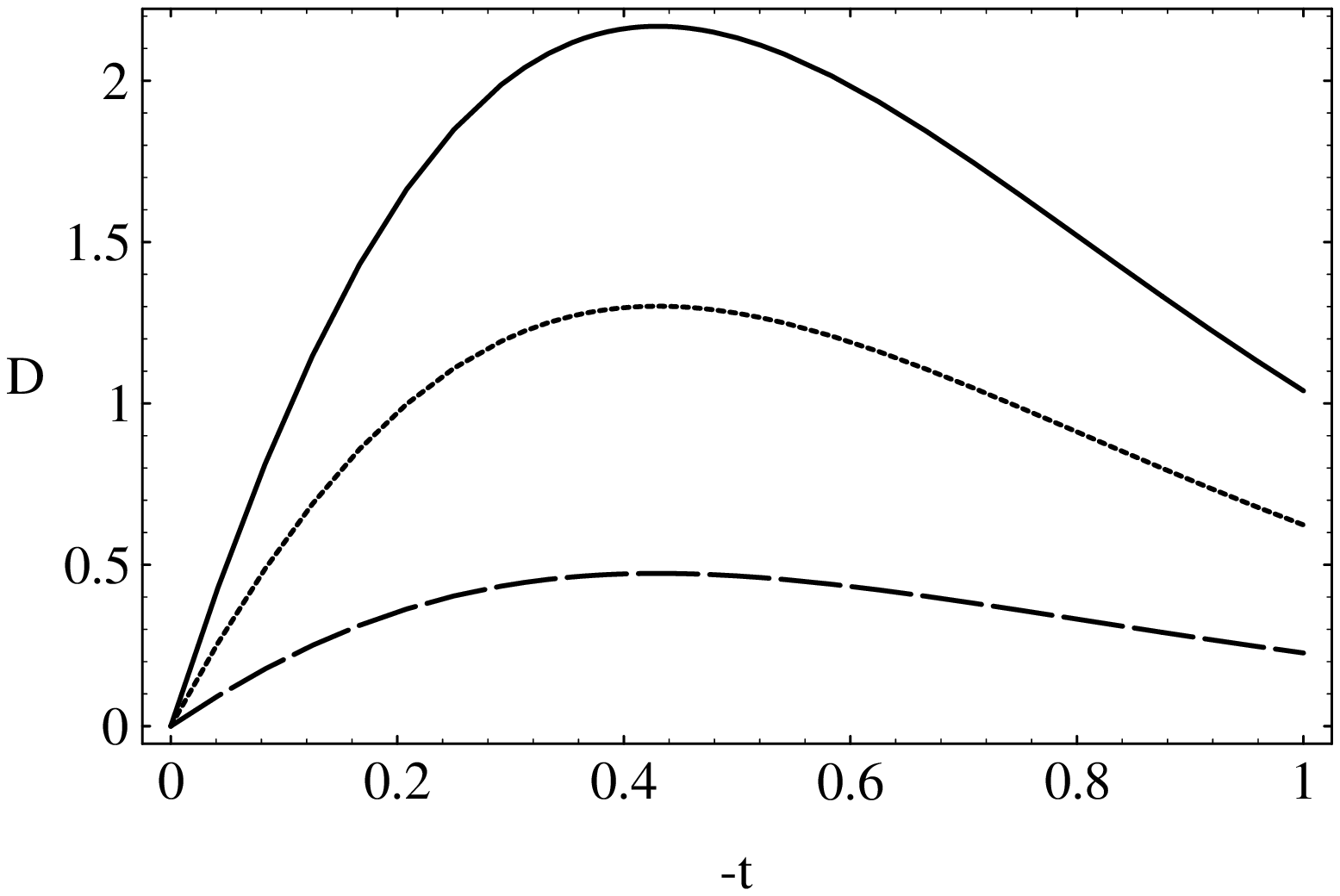,width=6.7cm}
\epsfig{file=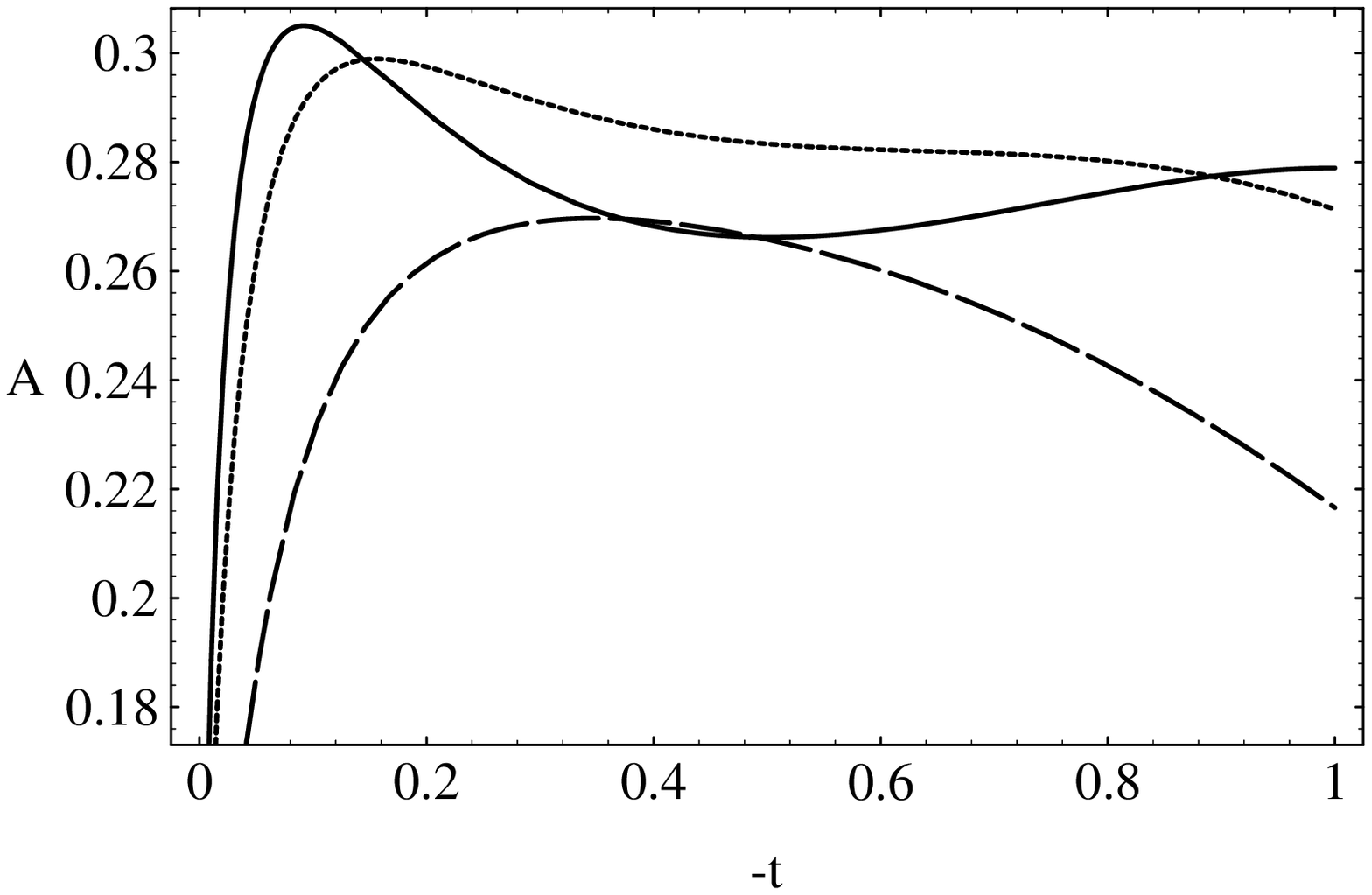,width=6.7cm}}
\vspace*{-4cm}
\caption{a) $D$ is plotted versus $-t$ for $x=10^{-4},10^{-3},10^{-2}$, 
$Q^2=12~\mbox{GeV}^2$, $B=5~\mbox{GeV}^{-2}$ and $y=0.4$. The solid curve is 
for $x=10^{-4}$, the dotted one for $x=10^{-3}$ and the dashed one for 
$x=10^{-2}$. b) The asymmetry $A$ is plotted versus $-t$ for $x=10^{-4}$ 
(solid curve), $x=10^{-3}$ (dotted curve) and $x=10^{-2}$ (dashed curve) again 
for $Q^2=12~\mbox{GeV}^2$, $B=5~\mbox{GeV}^{-2}$ and $y=0.4$.}
\label{tdep1}
\end{figure}
The complete cross section of exclusive photon production includes the 
Bethe-Heitler and DVCS process plus an interference term stemming from both processes. 
We find the cross section for DVCS, Bethe-Heitler and the interference term
to be
\bea
\frac{d\sigma_{DVCS}}{dxdyd|t|d\phi_r}&=&\frac{\pi\alpha^3s}{4R^2Q^6}
(1+(1-y)^2)e^{-B|t|}F^2_2(x,Q^2)(1+\eta^2),\\
\frac{d\sigma_{BH}}{dxdyd|t|d\phi_r}&=&
\frac{8\alpha^3sy^4}{\pi |t|Q^4(1-y)}\left[\frac{G^2_E(t)+
\frac{|t|}{4m_N^2}G^2_M(t)}{1+\frac{|t|}{4m_N^2}}\right ] cos^2(\phi_r),\\
\frac{d\sigma^{int}_{DVCS+BH}}{dxdyd|t|d\phi_r}&=& \pm 2\eta\frac{\alpha^3
sy^2F_2(x,Q^2)cos(\phi_r)e^{-B|t|/2}}{RQ^5}\sqrt{\frac{2(1+(1-y)^2)}{|t| 
(1-y)}}\nonumber\\
& &\times \left[\frac{G_E(t) + \frac{|t|}{4m_N^2}G_M(t)}{1+\frac{|t|}{4m_N^2}}
\right ].
\label{inter}
\eea
with $y=1-E'/E$, where $E'$ is the energy of the 
electron in the final state, $\phi_r=\phi_N - \phi_e$, 
where $\phi_N$ is the azimuthal angle of the final state proton with respect
to the reaction axis and $\phi_e$ is the azimuthal angle of the final state 
electron and where $G_E(t)$ and 
$G_M(t)$ are the electric and nucleon form factors. We describe them using 
the  well known dipole fit.
The + sign in the interference formula corresponds to an electron scattering 
off a proton and the - sign corresponds to the positron. 
The total cross section is then just the sum of the three terms in Eq.\ 
\ref{inter}.

At this point it is important to determine how large the Bethe-Heitler 
background is as compared to DVCS for HERA kinematics, hence, in the following 
discussion, we will estimate the ratio $D=<d\sigma_{DVCS+BH}/dxdydt>/
<d\sigma_{BH}/dxdydt>-1$ allowing a background comparison
with $<...>=\int_{0}^{2\pi}d\phi_r$. We find that $D > 1$ 
(see Fig.\ \ref{tdep1}a)) for $x=10^{-4}$, $Q^2=12~\mbox{GeV}^2$, $y=0.4$ and 
$-t\simeq 0.25~\mbox{GeV}^{-2}$.
It is convenient to illustrate the magnitude of the intereference term in 
the total cross section by considering the asymmetry for proton and either 
electron or positron to be in the same and opposite hemispheres $
A = [\int_{-\pi/2}^{\pi/2}d\phi_r d\sigma_{DVCS+BH} - \int^{3\pi/2}_{\pi/2}
d\phi_r d\sigma_{DVCS+BH}]/\int_{0}^{2\pi}d\phi_r d\sigma_{DVCS+BH}$. 
Fig.\ \ref{tdep1}b shows $A$ for the same kinematics as 
above and we find that the 
asymmetry is fairly sizeable already for small $t$ . Due to this fairly large 
asymmetry, one has a first chance to 
access nondiagonal parton distributions through it. 
Also, we find that $A$ is very sensitive to the energy dependence of the
scattering amplitude since it is proportional to $\eta$. As a result 
measurements of $A$ would provide a window to the energy behaviour of 
$F_{2N}(x,Q^2)$ beyond the energy range currently available at HERA.

Conclusion: In the above said we have shown that pQCD is applicable to 
exclusive photoproduction. We wrote down an equation for the imaginary part of 
the amplitude, which can be generalized to the complete amplitude at small $x$.
We also found that the imaginary part of the  amplitude
of the production of a real photon is larger than the one in 
the case of DIS in a broad range of $Q^2$.
We also found the same to be true for the full amplitude at
small $x$. We also make experimentally testable predictions for
the number of real photon events and suggest that the number of events are 
small but not too small to be detected at HERA. Finally, we demonstrated that 
first the Bethe-Heitler process does not overshadow the DVCS cross section 
and secondly that measuring the asymmetry $A$ at HERA would allow one to 
determine the real part of the DVCS amplitude, i.e.\ , gain a first 
experimental insight into nondiagonal parton distributions.


\begin{thebibliography}{99}
\bibitem{f5} This  effect will be taken into account in the actual numerical 
             calculation by incorporating the different evolution of diagonal 
             and nondiagonal distribution in the formula for the imaginary 
             part of the amplitude.
\end{thebibliography}
\end{document}